\documentclass[desactivate]{aa}
\usepackage[varg]{txfonts}
\usepackage{graphicx} 
\usepackage{natbib}
\bibpunct{(}{)}{;}{a}{}{,} 
\usepackage[colorlinks=true, 
            linkcolor=blue, 
            citecolor=blue, 
            filecolor=blue, 
            urlcolor=blue]{hyperref}

\usepackage{xcolor}
\usepackage{ulem}
\newcommand{\sunrise}{\textsc{Sunrise}}
\newcommand{\sunriseiii}{\textsc{Sunrise~iii}}

\makeatletter
\renewcommand*\aa@pageof{, page \thepage{} of \pageref*{LastPage}}
\makeatother

\title{Formation of chromospheric \ion{Fe}{I} lines in the near ultraviolet in 1D atmospheres}

\author{E. Harnes \inst{1,2} \and H. N. Smitha \inst{1} \and A. Korpi-Lagg \inst{1,3}
\and D. Przybylski \inst{1}
\and S. K. Solanki \inst{1}
}

\institute{Max Planck Institute for Solar System Research, Justus-von-Liebig-Weg 3, 37077 G{\"o}ttingen, Germany \\ \email{harnes@mps.mpg.de}
\and Georg-August-Universit{\"a}t G{\"o}ttingen, Friedrich-Hund-Platz 1, 37077 G{\"o}ttingen, Germany
\and Aalto University, Department of Computer Science, Konemiehentie 2, 02150 Espoo, Finland}

\date{Received  / Accepted }

\abstract{In the near ultraviolet (NUV) part of the solar spectrum,
there are several \ion{Fe}{I} lines with very broad profiles, typical of chromospheric lines. These lines are largely unexplored due to the lack of high-resolution data in this region. This changed with the successful \sunriseiii{} flight in 2024, when spectro-polarimetric data were recorded with high spatial, spectral, and temporal resolution covering a large variety of solar targets.
}
{The aim of this work is to investigate the formation of the lines and lay the groundwork for further studies.}
{We compute the spectrum of the lines at 358.12 nm, 371.99 nm, 406.36 nm, and 407.17 nm emerging from the standard 1D FAL-atmospheres using the non-local thermodynamic equilibrium (NLTE) radiative transfer code RH.}
{We find that the lines are affected by overionization in the wings, but have line cores strongly affected by scattering. The line cores form well into the chromosphere in the tested atmosphere models except the colder FALX model where the line core forms in the temperature minimum (which lies at traditional chromospheric heights). In the presence of a vertical magnetic field, the Stokes $V$ signal is multi-lobed. The profile can be decomposed into two broad photospheric lobes and two sharper lobes forming in the flanks of the chromospheric line core.}
{We have investigated the properties of four lines of Fe I that form in the lower chromosphere. The results provide a basic understanding of the formation of the lines, which will be useful for later analysis of formation in 3D magnetohydrodynamic simulations and an eventual investigation into their diagnostic potential.}


\begin{document}

\maketitle

\section{Introduction} \label{sec:introduction}

The solar spectrum is rich in \ion{Fe}{I} lines.
\citet{Athay_Lites_1972}, \citet{1973_Lites}, and \citet{1973_Lites_Brault} found that the \ion{Fe}{I} lines in the solar spectrum are sensitive to non-local thermodynamic equilibrium (NLTE) effects and in particular overionization. Excess UV photons ionize \ion{Fe}{I} in the photosphere from levels above 3-4\,eV. Due to the strong collisional and radiative coupling of the levels this effect spreads also to all other terms \citep{Rutten_1988, Shchukina_2001}.  Fewer neutral Fe atoms means that the absorption by \ion{Fe}{I} is comparatively weaker in the photosphere and around the temperature minimum, resulting in a weaker line than in local thermodynamic equilibrium (LTE). A second effect that certain lines may be affected by is scattering, mostly important when the collisional coupling becomes weaker in the chromosphere and upper photosphere \cite[]{Smitha_2020, Smitha_2021}.

In recent years, sophisticated atomic models of \ion{Fe}{I} have been implemented in solar and stellar modeling and abundance analysis, where the detailed \ion{Fe}{I} - \ion{Fe}{II} ionization balance is important, see for example \citet{Short_Hauschildt_2005, Shchukina_2005, Mashonkina_2011, Bergemann_2012, Lind_2017}. This development has come along with the increase in available atomic data on spectral lines, energy levels, photoionization and collisional cross-sections. Studies have also been carried out on how the solar \ion{Fe}{I} spectrum is dependent on 3D NLTE effects in increasingly realistic solar atmospheres \citep{Holzreuter_2013, Holzreuter_2015, Smitha_2021}.

The dense haze of spectral lines in the near ultraviolet (NUV) part of the solar spectrum harbors many \ion{Fe}{I} lines with strong and broad profiles, typical of chromospheric lines. 
The chromospheric iron lines are expected to provide information about the atmosphere over a large height interval, ranging from the photosphere to the chromosphere. In addition, the many surrounding lines can provide additional constraints on the photosphere and in some cases the lower chromosphere.

The diagnostic potential of these lines is largely unexplored due to a lack of high resolution observations in this spectral region. 
With the successful flight of \sunriseiii{} in July 2024 \citep{SunriseIII:2025}, we have for the first time captured high spatial, spectral and temporal resolution observations in some of the investigated Fe I lines using the Sunrise Ultraviolet Spectropolarimeter and Imager \citep[SUSI, ][]{Feller_2020, feller2025}. The observations range from quiet Sun to active regions .

In this paper, we focus on understanding the different physical processes influencing the formation of these chromsopheric \ion{Fe}{I} lines using simple one-dimensional atmospheres. 
Their formation in realistic 3D MHD simulations and comparison with \sunriseiii{} observations will be addressed in follow-up work. 

The structure of this paper is as follows; In Sect. \ref{sec:methods} we introduce the methods used to obtain the spectra, in Sect. \ref{sec:line_profiles} we present the resulting spectra, in Sect. \ref{sec:nlte_effects} we discuss the NLTE effects at play and in Sect. \ref{sec:opacity_fudge} we briefly discuss the effects of the treatment of the UV continuum. A summary is given in Sect. \ref{sec:conclusion}.

\section{Spectral Synthesis} \label{sec:methods}

We synthesized the spectral profiles using the RH radiative transfer code \citep{uitenbroek_multilevel_2001}. The specific lines studied here in detail were selected based on the \sunriseiii{} SUSI observation programs, which included time series of spectral regions covering the 358.12\,nm, 406.36\,nm, and 407.17\,nm lines. During the \sunrise{} flight, a scan covering large parts of the spectral window of SUSI (309\,nm to 417\,nm) was done. This scan briefly covered the line at 371.99\,nm, which may be particularly interesting since it is a resonance line and its core is relatively free from nearby blend lines. We also discuss the formation properties of this line. 

We initially started with an iron model atom with 33 levels\footnote{Available here: \url{https://github.com/han-uitenbroek/RH/blob/master/Atoms/Fe33.atom}}, and updated level energies from the National Institute of Standards and Technology (NIST) atomic lines database \citep{NIST_ASD}. For line broadening by collisions with neutral hydrogen atoms, we use the results of \citet{Anstee_1995} for s-p and p-s transitions.
Since the 33-level model atom did not include all the lines of our interest, the iron atom model had to be significantly improved. This included splitting the term 3d$^6$ 4s 4p\,$^5\mathrm{F}^\circ$ into individual levels and updating the oscillator strengths and radiative broadening of the lines with recent values available in the NIST database and the Vienna Atomic Line Database \citep[VALD, ][]{Ryabchikova_2015}. The 3d$^7$\,4p\,\,$^5\mathrm{G}^\circ$, 3d$^7$\,4s\,\,$^5\mathrm{F}$, 3d$^7$\,4s\,\,$^3\mathrm{F}$, and 3d$^7$\,4p\,\,$^3\mathrm{F}^\circ$ levels were added. From each new level added, the list of bound-free photoionization rates was expanded assuming hydrogenic approximation, we added collisional rates to all the other levels, and we included a selection of line transitions to other levels in the iron model using the NIST database. When available, explicit photoionization rates from \cite{1997_Bautista} and the NORAD database \citep{Nahar_2020, Nahar_2024} were included in the atom model. The result was a 52-level model atom with 167 bound-bound transitions, including the lines at 358.12, 371.99, 406.36, and 407.17\,nm. Some key parameters for these lines are shown in Table \ref{tab:lines-table} along with parameters of a few similar lines of \ion{Fe}{I} in the NUV. 
These lines are strong and broad, and the lower levels of their transitions are from the lowest terms of \ion{Fe}{I}, 3d$^6$\,4s$^2$\,\,$^5$D, 3d$^7$\,4s\,\,$^5$F, and 3d$^7$\,4s\,\,$^3$F. It should be noted that there are other broad \ion{Fe}{I} lines in this region that can also be studied in the SUSI full spectral scans (between 309\,nm and 417\,nm), outside the selection made for this paper. See e.g. identified lines with large equivalent width in \citet{1966_Moore}.

For our tests we used the FAL semi-empirical model atmospheres A, C, F, and P \citep{1993_Fontenla} and FALX \citep{1995_Avrett}. The models encompass quiet sun (A, C, F) and plage (P), while the FALX model was constructed using the CO lines and represents the quietest inter-network. Their temperature structures can be seen in Fig. \ref{fig:FAL-temp}. These atmospheres provide some variation in parameters that can reveal the sensitivity of the lines to the temperature structure. To investigate the sensitivity to the magnetic field, we also computed the Stokes $V$ profiles with an imposed vertical, height-independent magnetic field with strengths of 0\,G, 200\,G, 1\,kG, and 2.5\,kG in the FALC atmosphere. We used the field-free method of \citet{Rees_1969}, which means we assume that the effect of the magnetic field on the level populations is negligible. As we use height-independent magnetic fields and the lines are broad and are only weakly split at the magnetic field strengths we test, the field-free approximation is applicable \citep{1977_Auer_Heasley_House, 1996_Bruls_Trujillo}.

As the \ion{Fe}{I} level populations and the \ion{Fe}{I}/\ion{Fe}{II} ionization balance are highly sensitive to the UV radiation in the photosphere, it is important to also properly model the UV continuum \citep{1973_Lites, Rutten_1988, Shchukina_2001}. In addition to the continuous opacity sources from the atoms H, He, C, N, O, Na, Mg, Al, Si, S, Ca, which are all treated in LTE, we utilize the UV opacity fudge factors of \citet{1993_Bruls} to account for the UV line haze and the missing opacity. These opacity fudge factors were based on empirical fitting of the calculated UV continuum to disk center observations listed in \citet[][Table 10]{1976_Vernazza}. 
A more detailed treatment of the UV continuum would involve simultaneous NLTE computations of other minority species such as Si, Al, and Mg that contribute significantly with photoionization opacity in the region and fix the UV photon density available for iron photoionization, self-consistent electron densities, and lastly including the millions of lines making up the UV line haze \citet{1976_Vernazza, Rutten_2019, Rutten_2021, Smitha_2023}. We discuss the effects of the fudge factors in Sect. \ref{sec:opacity_fudge}.

\begin{table*}[htb!]
    \caption{Atomic data for a collection of \ion{Fe}{I} lines in the NUV.}
    \label{tab:lines-table}
    \centering
    \begin{tabular}{l l l l l l l r l}
        \hline
        $\lambda$ $[$nm$]$ & lower level & lower level & E$_\mathrm{l}$ [eV]& upper level & upper level & E$_\mathrm{u}$ [eV] & log(gf) & eff. Landé\\
        & configuration & term & & configuration & term & & & g-factor \\
        \hline
        271.90          & 3d$^6$ 4s$^2$ & $^5$D$_4$          & 0.00   & 3d$^6$ 4s 4p & $^5$P$^{\circ}_3$      & 4.56 &  0.042          & 1.26 \\
        296.69          & 3d$^6$ 4s$^2$ & $^5$D$_4$          & 0.00   & 3d$^7$ 4p & $^5$F$^{\circ}_5$      & 4.18 & -0.404          & 1.25 \\
        358.12 & 3d$^7$ 4s & $^5$F$_5$ & 0.86 & 3d$^7$ 4p & $^5$G$^{\circ}_6$ & 4.32 & 0.406  & 1.16 \\
        371.99 & 3d$^6$ 4s$^2$ & $^5$D$_4$  & 0.00 & 3d$^6$ 4s 4p & $^5$F$^{\circ}_5$ & 3.33 & -0.432 & 1.20 \\
        373.49          & 3d$^7$ 4s & $^5$F$_5$          & 0.86   & 3d$^7$ 4p & $^5$F$^{\circ}_5$      & 4.18 & 0.317           &  1.41 \\
        373.71          & 3d$^6$ 4s$^2$ & $^5$D$_3$ 	       & 0.05   & 3d$^6$ 4s 4p & $^5$F$^{\circ}_4$       & 3.37 & -0.574          & 1.14 \\ 
        382.04          & 3d$^7$ 4s & $^5$F$_5$          & 0.86   & 3d$^7$ 4p & $^5$D$^{\circ}_4$      & 4.10 & 0.119           & 1.21 \\
        385.99          & 3d$^6$ 4s$^2$ & $^5$D$_4$          & 0.00   & 3d$^6$ 4s 4p & $^5$D$^{\circ}_4$      & 3.21 & -0.710          & 1.50 \\
        404.58          & 3d$^7$ 4s & $^3$F$_4$          & 1.48   & 3d$^7$ 4p & $^3$F$^{\circ}_4$             & 4.55 & 0.280           & 1.25 \\
        406.36 & 3d$^7$ 4s & $^3$F$_3$ & 1.56   & 3d$^7$ 4p & $^3$F$^{\circ}_3$    & 4.61 & 0.062  & 1.09  \\ 
        407.17 & 3d$^7$ 4s & $^3$F$_2$ & 1.61   & 3d$^7$ 4p & $^3$F$^{\circ}_2$    & 4.65 & -0.022 & 0.68  \\
        \hline
    \end{tabular}
    \tablefoot{These lines appear strong and broad in the solar spectrum and are likely chromospheric. Energy levels and log(gf) were taken from NIST. Effective Landé g-factors were extracted from the VALD database. The ${\circ}$ symbol indicates odd-parity terms.}
\end{table*}

\section{Line profiles in the FAL atmospheres} \label{sec:line_profiles}

\begin{figure}[htb]
    \centering
    \includegraphics[width=\linewidth]{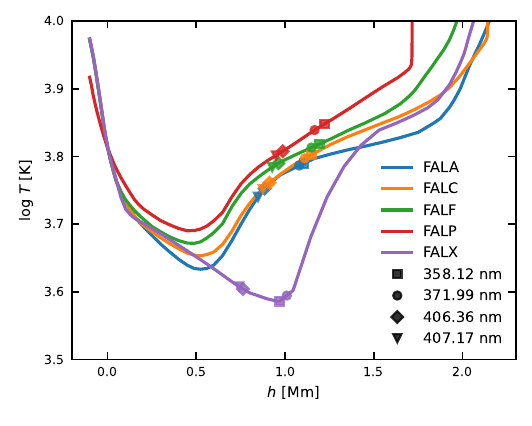}
    \caption{Temperature structure of the FAL models. The markers show the height where $\tau_{\nu} = 1$ for each line core.}
    \label{fig:FAL-temp}
\end{figure}

\begin{figure*}[htb!]
        \centering
        \includegraphics[width=\linewidth]{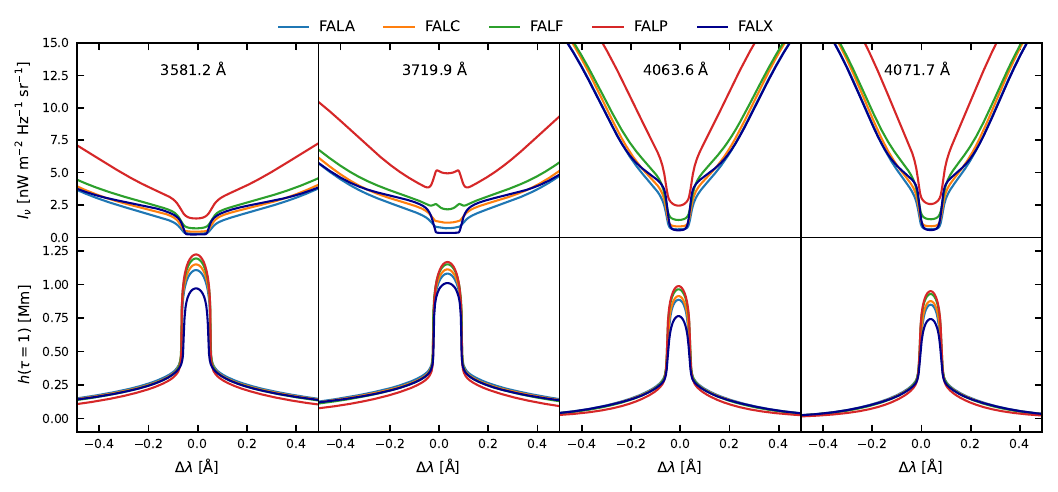}
        \caption{Top row: Emergent intensity in the FAL atmospheres near the cores of the four \ion{Fe}{I} lines of interest. Bottom row: Height where $\tau_{\nu}=1$ in the respective atmospheres and lines.}
        \label{fig:FAL_profiles}
\end{figure*}

Figure \ref{fig:FAL-temp} shows the temperature structure of the models along with the locations where the optical depth at the line core wavelengths, of the four lines computed here, is one. The emergent intensity of the four selected lines are shown in Fig. \ref{fig:FAL_profiles} along with the height where the optical depth at the corresponding wavelength is equal to unity as an indication of the formation height of the lines. The wings form in the photosphere. The line cores form, for all atmospheres except FALX, well above the temperature minimum (located at approximately 0.5\,Mm), i.e in the chromosphere of these atmosphere models. 
The FALX model has a different temperature stratification, with the temperature minimum being much cooler than the others and is located at approximately 1\,Mm. We find that the height of formation is correlated with the temperature stratification of the atmosphere, with the lines forming higher in the warmer atmospheres. The 406.36\,nm and 407.17\,nm lines form a few hundred km below the 358.12\,nm and 372\,nm lines in all the five FAL models.

All the FAL atmospheres display very similar profiles in the wings except for the warmer FALP atmosphere, which has significantly weaker wings. The core of the lines in the FALX atmosphere are stronger than in the other atmospheres. 
This is caused by the lower temperature of the FALX atmosphere compared to the other atmospheres at the formation height of the line cores.

The 372\,nm line stands out compared to the 358.12, 406.36, and 407.17\,nm lines in that it has two emission peaks in the line core in the warm FALP and FALF atmospheres. The 372\,nm line is a resonance line, with the ground level being the ground state of \ion{Fe}{I}, the level 3d$^6$\,4s$^2$\,\,$^5$D$_4$. This is discussed further in Sect. \ref{sec:nlte_effects}.

\begin{figure*}[htb!]
    \centering
    \includegraphics[width=\linewidth]{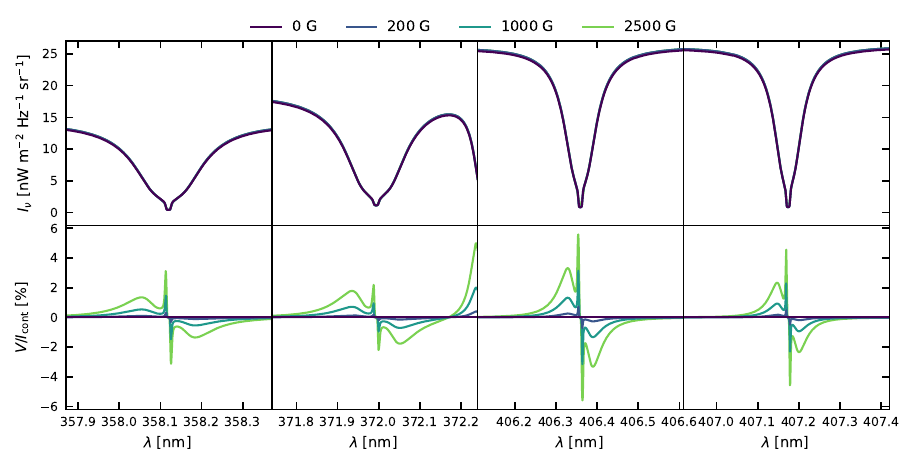}
    \caption{Stokes $I$ (upper row) and Stokes $V$ (lower row) for the FALC atmosphere with vertical, height-independent magnetic field strengths of 0\,G, 200\,G, 1\,kG and 2.5\,kG imposed.
    }
    \label{fig:stokes}
\end{figure*}

The Stokes $V$ profiles computed with an imposed vertical magnetic field of constant strength are shown in Fig.~\ref{fig:stokes}. The top row shows the Stokes-I profiles and the bottom row the Stokes-V profiles. The red wing of the 372\,nm line is blended with the 372.3 nm \ion{Fe}{i} line, which is also present in the model atom. We find that the lines produce double-lobed Stokes $V$ profiles caused by the flattening of the inner line wings and the steep core. The outer lobes are photospheric, and the inner, narrow lobes form in the flanks of the chromospheric line core. Surrounding lines, e.g. \ion{Fe}{I} in the red wing of the 372\,nm line at 372.26\,nm, are narrower and produce stronger Stokes-V signal. Although the lines at 358.12\,nm, 372\,nm, and 406.36\,nm have roughly similar effective Landé factors, at 2.5\,kG the narrower line at 406.36\,nm produces a central peak of almost $6\%$ while the 372\,nm line produces a signal closer to 2$\%$. We note that in a realistic atmosphere, the field strength will decrease with height, and the ratio of the amplitudes of the inner lobes to the outer lobes will also be smaller.

\section{NLTE effects} \label{sec:nlte_effects}

\begin{figure}[htb!]
    \centering
    \includegraphics[width=\linewidth]{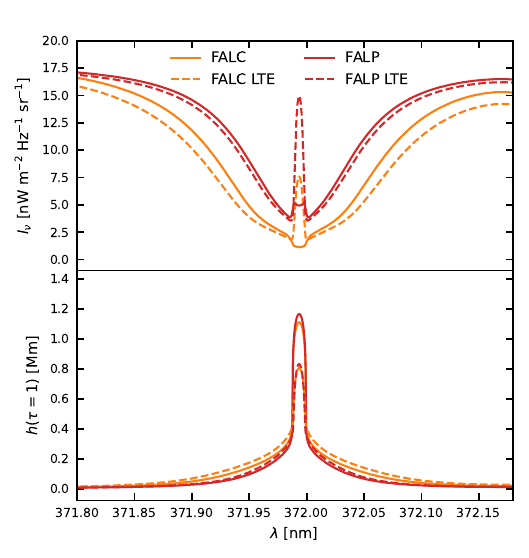}
    \caption{Intensity (upper row) and height where $\tau_{\nu}=1$ (lower row) in FALC (orange) and FALP (red) atmospheres. The solid lines are computed in NLTE, and the dashed lines in LTE.}
    \label{fig:FAL_profiles_focus}
\end{figure}

The effect of an NLTE vs. LTE treatment of the 372\,nm \ion{Fe}{I} line is shown in Fig. \ref{fig:FAL_profiles_focus} for the FALC and FALP atmospheres.  We find that the lines are affected by NLTE over their entire profile, and it is clear that an LTE assumption fails. In the wings, the lines are weaker in NLTE and stronger in LTE. On the other hand, the LTE assumption leads to an emission core while the core is in absorption in an NLTE treatment. The difference in intensity in the wings is larger in the FALC atmosphere than in the warmer FALP atmosphere, while in the core the opposite is true. It should be noted that the FALP atmosphere is treated with the same fudge factors as the FALC atmosphere, and a proper consideration of a warmer atmosphere would need NLTE modeling of the UV opacities, see Sect. \ref{sec:opacity_fudge}. 

\begin{figure}[htb!]
    \includegraphics[width=\linewidth]{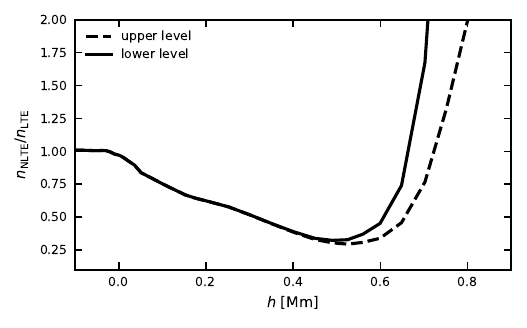}
    \caption{Departure coefficients for the 372\,nm line transition in the FALC atmosphere. The solid line is for the upper level, and the dashed line is for the lower level.}
    \label{fig:departure_coefficients}
\end{figure}

The difference in LTE and NLTE in the wings is caused by the overionization of neutral iron in the photosphere. We use the definition of the departure coefficients as the ratio of the NLTE populations of a given level to the ratio of the LTE populations of the same level, 
\begin{equation}
    \beta_i = \frac{n_{i,\mathrm{NLTE}}}{n_{i,\mathrm{LTE}}},
\end{equation}
following \citet{Wjibenga_Zwaan_1972}.
Figure \ref{fig:departure_coefficients} shows the departure coefficients of the upper and lower levels of the 372\,nm line for the FALC model. In the photosphere the NLTE populations are lower than the LTE populations, but the upper and lower level departure coefficients are equal. As long as $e^{h\nu/kT}>>\beta_u/\beta_l$ the line source function, $S_{\nu}$, is 
\begin{equation}
    S_{\nu} \approx  \frac{\beta_u}{\beta_l}B_{\nu}(T),
\end{equation} 
where $B(T)$ is the Planck function (follows from Eq. 1.70 and 1.79 in \citet{Rybicki_1986}). It follows that if the departure coefficients for the upper and lower levels are equal, then the source function will follow the Planck function, even if the populations in each level are departing from the LTE populations. The reduction in line strength comes from the smaller opacity caused by the deficit in \ion{Fe}{i}. The line extinction coefficient can, when $\beta_u = \beta_l$, be expressed as
\begin{equation}
    \alpha_{\nu} = \beta_l \alpha_{\nu, \mathrm{LTE}}
\end{equation} 
where $\alpha_{\nu,\mathrm{LTE}}$ is the line extinction coefficient for LTE populations (follows from Eq. 1.78 in \citet{Rybicki_1986}). So similarly to the ratio of the NLTE to LTE source functions, we can read the ratio of the NLTE to LTE line extinction coefficients from Fig. \ref{fig:departure_coefficients}. In the regions where the wings form, roughly below 0.4 Mm, this ratio is less than 1. However, the departure coefficients approach the value of 1 in the deepest part of the atmosphere as the level populations approach LTE.

\begin{figure*}
    \resizebox{\hsize}{!}{\includegraphics{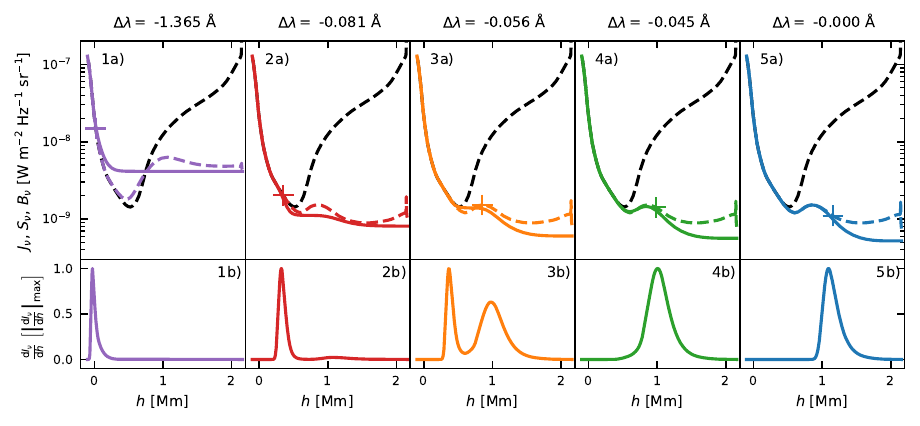}}
    \caption{Upper row: Height stratification of the source function (colored dashed lines), angle-averaged intensity (colored solid lines), and the Planck function (black dashed line) in the FALC atmosphere. The crosses mark the locations where $\tau_{\nu}=1$. The intensity of the four wavelengths points closest to the rest wavelength is marked with crosses of corresponding color in Fig.~\ref{fig:FALC_FALP_cores}. The purple line is taken from the wing at 371.85\,nm. Second row: Contribution functions at wavelengths corresponding to the upper row.}
    \label{fig:FALC_JSB}
\end{figure*}

The overionization effect is important throughout the wings of the lines. The contribution function, the integrand of the formal solution to the radiative transfer equation,
\begin{equation}
    \frac{\mathrm{d}I_{\nu}}{\mathrm{d}h} = S_{\nu}(h)e^{-\tau}\frac{\mathrm{d}\tau}{\mathrm{d}h}
\end{equation} is shown for a point in the wing of the 372 nm line at 371.85\,nm in panel 1b) in Fig. \ref{fig:FALC_JSB}. 
At this wavelength, the formation height of the line is spread out in a distribution around 0 Mm. When comparing this with the departure coefficients for the FALC atmosphere in Fig. \ref{fig:departure_coefficients}, it is clear that even far out in the wings, the ionization balance contributes to the line strength. \citet{1973_Lites} defined the far wings as the part of the line wings forming in LTE, and the inner wings as the part of the wings forming above the height where the \ion{Fe}{i}-\ion{Fe}{ii} balance starts to depart from LTE. \citet{1973_Lites} found that this transition was usually located at the wavelength where the intensity was around $25\%$ of the continuum. It should be noted that the continuum intensity changes somewhat from LTE to NLTE in this region, due to the photoionization of \ion{Fe}{I}. We find that the change in the continuum is not sufficient enough to explain the significant differences in the LTE and NLTE wings even further out than at $25\%$ of the continuum intensity. This is explained by the low height at which the departure coefficients start to deviate from 1 and the spread in the contribution function. 

\begin{figure*}
    \resizebox{\hsize}{!}{\includegraphics{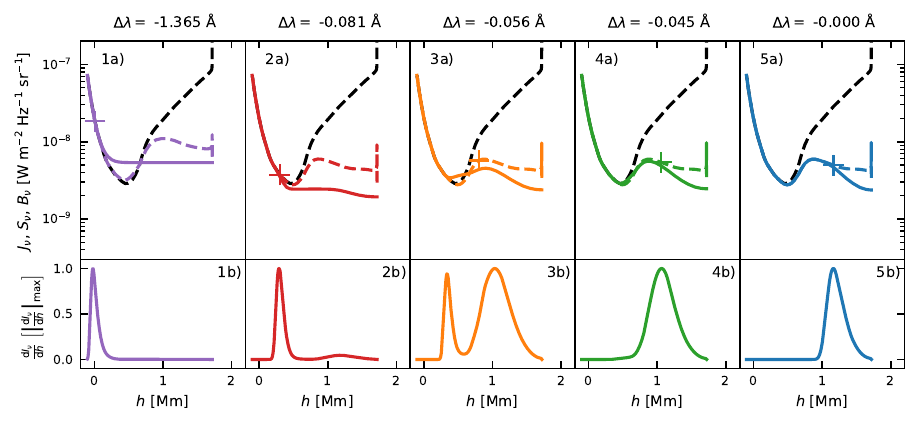}}
    \caption{Same as Fig.~\ref{fig:FALC_JSB}, but for the FALP atmosphere.}
    \label{fig:FALP_JSB}
\end{figure*}

\begin{figure}[htb!]
    \resizebox{\hsize}{!}{\includegraphics{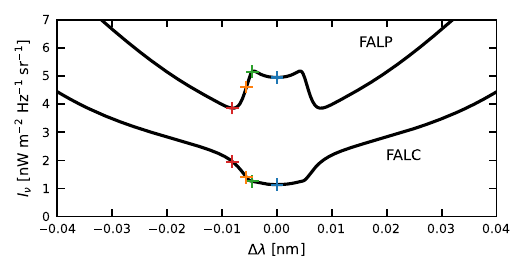}}
    \caption{The line core of the 372\,nm line with the FALC and FALP atmospheres. The crosses mark the wavelengths where the height stratifications of the source functions are plotted in Fig. \ref{fig:FALC_JSB} and Fig. \ref{fig:FALP_JSB}.}
    \label{fig:FALC_FALP_cores}
\end{figure}

The cores of the lines are scattering-dominated. Figure \ref{fig:FALC_JSB} and Fig. \ref{fig:FALP_JSB} col. 2-5 show how the source function for four sample points near the core is affected by scattering in the chromosphere. The locations of the points are shown in Fig. \ref{fig:FALC_FALP_cores}. The source function, $S$, decouples from the Planck function, $B$, and follows the angle-averaged intensity, $J$, instead. This behavior is common among the four lines studied here, but the 372\,nm line stands out with its behavior in the FALP atmosphere. 

The line core of the 372\,nm resonance line shows a double-peaked emission profile in the core for the FALP and FALF atmospheres. A comparison between the  line cores of the 372\,nm line formed in FALP and FALC atmospheres is shown in Fig. \ref{fig:FALC_FALP_cores}. 
At the wavelength of the red cross, the corresponding contribution functions in panels 2b) in Fig. \ref{fig:FALC_JSB} and Fig. \ref{fig:FALP_JSB} show that the height where $\tau = 1$ is below the temperature minimum and in the region where the source function still follows the Planck function. When looking at the contribution functions, there is a jump in the formation height between the red cross (col. 2 in Fig. \ref{fig:FALC_JSB} and Fig. \ref{fig:FALP_JSB}) and the green cross (col. 4 in Fig. \ref{fig:FALC_JSB} and Fig. \ref{fig:FALP_JSB}). This is also visible when looking at a point taken from the flank of the core, see col. 3 in Fig. \ref{fig:FALC_JSB} and Fig. \ref{fig:FALP_JSB}. At the green cross, the line forms above 1\,Mm in the FALP atmosphere, and the source function is here still sensitive to the temperature increase, which causes the relative increase in emerging intensity at this wavelength. At the rest wavelength, marked by the blue cross (col. 5 in Fig. \ref{fig:FALC_JSB} and Fig. \ref{fig:FALP_JSB}), the formation height is greater and the source function is  more decoupled from the Planck function.

\begin{figure}[htb!]
    \resizebox{\hsize}{!}{\includegraphics{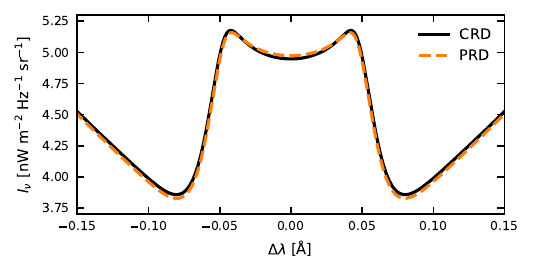}}
    \caption{The line core computed with PRD (orange dashed line) vs. CRD (black solid line) in the FALP atmosphere for the 372\,nm resonance line.}
    \label{fig:FALP_PRD}
\end{figure}

As the 372\,nm line is producing a double-peaked line core reversal similar to \ion{Ca}{II} H \& K lines \cite[see e.g.][]{1974_Vardavas, 1975_Shine, Neckel_1999}, we also checked if it would be sensitive to partial redistribution (PRD) or if the assumption of complete redistribution (CRD) would be a good approximation. In Fig. \ref{fig:FALP_PRD} a comparison of the line core is shown under the two different assumptions. The differences are minimal,  with the PRD profile being only slightly deeper at the location of the reversal and slightly shallower at the central wavelength. This implies that it will likely be sufficient to treat this line with the CRD assumption, but testing in 3D MHD cubes is needed.

\section{The UV continuum} \label{sec:opacity_fudge}

\begin{figure}[htb]
    \centering
    \resizebox{\hsize}{!}{\includegraphics{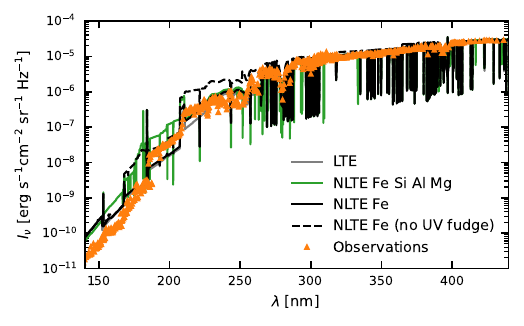}}
    \caption{Comparison of the synthesized broadband UV spectra against the observations (orange triangles) used by \citet{1993_Bruls} to make the opacity fudge factors. The black and green solid lines are computed with Fe treated in NLTE, and Fe, Si, Al and Mg treated in NLTE, respectively. The dashed black line is computed with Fe in NLTE after removing the UV fudge factors. The LTE equivalent from Fig. \ref{fig:hamburg-comparison} is not plotted here as the UV continuum is not important for the formation of the line wings in LTE.}
    \label{fig:uv-continuum}
\end{figure}

\begin{figure}[htb]
    \centering
    \resizebox{\hsize}{!}{\includegraphics{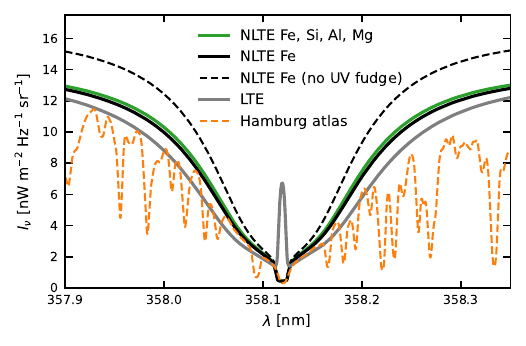}}
    \caption{Comparison of the emerging line profile for the 358.1\,nm line synthesized with different assumptions affecting the ionization balance of Fe, against the Hamburg atlas. The orange dashed line is the Hamburg atlas. The solid gray line is the LTE synthesis, and the solid black line is synthesized with Fe treated in NLTE. Removing the UV fudge factors from the NLTE Fe synthesis results in the dashed black  line. Including the fudge factors in addition to treating Si, Al, and Mg in NLTE (hence modifying the UV continuum) results in the green line. }
    \label{fig:hamburg-comparison}
\end{figure}

\begin{figure*}[htb]
    \centering
    \includegraphics[width=\linewidth]{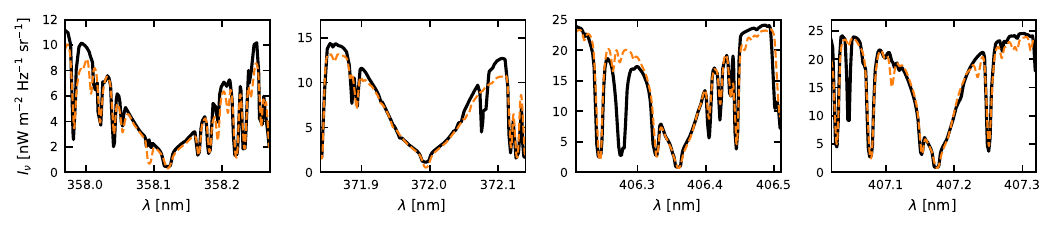}
    \caption{Comparison of the results from the FALC atmosphere (solid black line) with the Hamburg atlas (dashed orange line). In the synthesis, only iron has been treated in NLTE, with the surrounding lines from the Kurucz line list treated in LTE.}
    \label{fig:hamburg-atlas}
\end{figure*}

The overionization effect discussed in Sect. \ref{sec:nlte_effects} is driven by the UV radiation and the bound-free edges of \ion{Fe}{I} that can be found there. The UV part of the spectrum is affected by the bound-free edges of many atoms, some of which are also subject to overionization due to the UV radiation, in particular \ion{Si}{I}, \ion{Mg}{I}, and \ion{Al}{I}. On top of the NLTE effects from the overionization of these atoms, the dense UV line haze plays a part in reducing the NUV intensity. To properly deal with this, the NLTE effects of both the continuous opacity sources and the millions of lines would need to be treated at the same time \citep{Rutten_2019}. This requires not only an infeasible amount of computational power to solve, but also accurate atomic data of both spectral lines and photoionization rates.  

There are several methods to deal with this problem with considerable simplifications. One possible way to simplify this problem is to account for the UV line haze using opacity distribution functions (see e.g. \citet{Busa_2001, Shapiro_2010}). Another way would be treat the UV line haze as continuum and use multiplicative factors to enhance the continuum opacity itself. The latter is what we have done in this work using the opacity fudge factors available in the RH code that were determined by \citet{1993_Bruls}. 

The question stands whether the opacity fudge factors from \citet{1993_Bruls} are appropriate for use with our opacity setup. As discussed in Sect. \ref{sec:nlte_effects}, the line wings are highly sensitive to the \ion{Fe}{I} overionization caused by excess UV radiation. To check the robustness of this simplification we compare our synthesized UV spectrum to the observed spectrum used by \citet{1993_Bruls} in Fig. \ref{fig:uv-continuum}. We find that, treating only \ion{Fe}{I} in NLTE results in a spectrum that for most wavelengths above 300\,nm and in areas around 250\,nm, 210\,nm, and below 180\,nm is somewhat too high. The opposite can be said for the interval around 200\,nm, but this deficit is similar to the original result of \citet{1993_Bruls}. 

Figure \ref{fig:hamburg-comparison} shows the 358.12 nm line vs. the Hamburg FTS atlas, given the various assumptions for the UV continuum shown in Fig. \ref{fig:uv-continuum}.
Combining the fudge factors with NLTE treatment of Si, Al, and Mg causes excess ionization of \ion{Fe}{I}, as the ionization of \ion{Si}{I}, \ion{Al}{I}, and \ion{Mg}{I} lifts the UV intensity \citep{Smitha_2023} and acts in opposition to the opacity fudge factors. This effect is largest around 200\,nm where the intensity is up to an order of magnitude greater. \ion{Fe}{I} however, is a major contributor to the opacity in the region around 250\,nm where this effect is smaller. The effect this has on the 358.12\,nm spectral line can be seen in Fig. \ref{fig:hamburg-comparison}. The wings of the line are slightly weaker when \ion{Si}{I}, \ion{Al}{I}, and \ion{Mg}{I} are also treated in NLTE. In addition, we tested this effect using the hydrogenic-like approximation of the original RH \ion{Fe}{I} model atom, where the photoionization cross-section is assumed to be inversely proportional to the frequency cubed. See \citet[][p. 96-105]{1978_Mihalas}. The much smaller photoionization cross-sections of \ion{Fe}{I} caused the wings of the lines to be much more sensitive to the treatment of \ion{Si}{I}, \ion{Al}{I}, and \ion{Mg}{I} in NLTE.
It is apparent when comparing our results to the Hamburg FTS atlas \citep{Neckel_1999} in Fig. \ref{fig:hamburg-comparison} that the overionization of the wings is an important part of reproducing observations. The LTE line profile has stronger inner wings compared to the Hamburg FTS atlas. 

To evaluate the model with observational data we add the surrounding lines from the Kurucz line list\footnote{Available here: \href{http://kurucz.harvard.edu/linelists/gfnew/}{http://kurucz.harvard.edu/linelists/gfnew/}} \citep{Kurucz_2018} 
and treat them in LTE. Figure \ref{fig:hamburg-atlas} shows the resulting intensity profile compared with the Hamburg FTS atlas. Our synthesis fits in general well with the atlas. The line cores of the iron lines fit well except for the 372\,nm line, which may be more dependent on the chromospheric temperature structure of the FALC model than the other lines. Many surrounding blend lines are poorly fitted, which is likely due to a combination of poor atomic data and the fact that they are treated in LTE only. We also do not include molecular blends, which could also explain some of the missing lines. 

Our tests suggest that the cores of the chromospheric \ion{Fe}{I} lines considered in this paper, are not very sensitive to the treatment of the UV continuum, nor the replacement of the hydrogenic approximation with the more accurate photoionization cross sections from \citet{1997_Bautista}. This aligns with the findings of  \citet{1973_Lites}, who discusses that the most important factors influencing the depth of the line cores are the collisional cross-sections and the atmospheric model through the temperature structure and electron density.

\section{Conclusions} \label{sec:conclusion}

This paper presents a detailed analysis of the formation of strong and broad \ion{Fe}{I} lines in the NUV, which are formed well into the chromosphere, and present a stark contrast to the widely used photospheric \ion{Fe}{I} in the visible and infrared wavelengths. This study was carried out as a preparatory investigation, which in the future will enable us to exploit the diagnostic potential of these interesting lines, which will help the analysis of the high spectral, spatial, and temporal observations of the recent, highly successful, \sunriseiii{} flight. 
To this end, as a first step, we synthesized the spectra for a selection of NUV chromospheric iron lines in NLTE using the one-dimensional FAL atmospheres. These lines are very broad and deep, typical of chromospheric lines such as the \ion{Ca}{II} 854.2\,nm line. 
We found that the wings of these neutral iron lines, which sample the photosphere, are affected by UV overionization. The cores forming at chromospheric heights are mainly scattering dominated. This is in accordance with the findings of \citet{1973_Lites} who investigated the formation of some of the chromospheric iron lines using the HSRA \citep{Gingerich_1971} atmosphere. 

We also computed the Stokes $V$ profiles by adding a constant vertical magnetic field of different strengths. The Stokes $V$ profiles of all four lines show an interesting four-lobed structure, with two outer lobes sampling the photosphere and two inner lobes sampling the flanks of the line core, which can be chromospheric. The signal in the strongest lobe is a few $\%$ even when the field strength is a couple of kG. A constant vertical field represents an ideal scenario, and in reality, the Sun's magnetic field is quite complex \citep[see e.g.][]{2019_Bellot_Rubio_Orozco_Suarez}. The ratio of the inner to outer lobes of the Stokes $V$ profiles, as well as their shapes, will be different in the more complex, height-dependent magnetic fields of the Sun, and further study of the lines is needed with realistic atmospheres and magnetic field stratifications. The four lines studied here have weaker $V/I_c$ signal than the \ion{Ca}{II} 854.2\,nm line. 

We find that the treatment of the UV continuum is a significant factor in shaping the wings of the investigated lines, and the inclusion of transitions to other levels of the Fe atom affects the line cores. Despite the  simplifications made in terms of the size of the iron atom model, treatment of UV line haze, and treatment of the electron densities,
our synthesized spectra from the FAL models provide a reasonable match to the spatially and temporally averaged line profiles in the Hamburg FTS atlas. Our next step is to extend this analysis by investigating the formation of these interesting lines in realistic 3D magnetohydrodynamic simulations carried out with the chromospheric extension of the MURaM code \citep{Przybylski_2022} and to explore their diagnostic capabilities in complement to other chromospheric lines such as the \ion{Ca}{II} H \& K \citep{2018_Bjoergen, 2024_Pandit_Wedemeyer_Carlsson}, \ion{Ca}{II} 854.2\,nm \citep{2016_Quintero_Noda}, and \ion{Mg}{I} b$_2$ \citep{2025_Siu-Tapia_a, 2025_Siu-Tapia_b} observed by \sunriseiii{}.

\begin{acknowledgements}
    We thank Du{\v s}an Vukadinovi{\'c} for help and discussions while developing the analysis. We thank Hans-Peter Doerr for providing the digital version of the Hamburg Atlas and Jo Bruls for providing information about the UV observations. This work was supported by the International Max-Planck Research School (IMPRS) for Solar System Science at the University of Göttingen. This project has received funding from the European Research Council (ERC) under the European Union's Horizon 2020 research and innovation programme (grant agreement No. 101097844 — project WINSUN).  This work was supported by the Deutsches Zentrum f{\"u}r Luft und Raumfahrt (DLR; German Aerospace Center) by grant DLR-FKZ 50OU2201. We gratefully acknowledge the computational resources provided by the Cobra and Raven supercomputer systems of the Max Planck Computing and Data Facility (MPCDF) in Garching, Germany. This research has made use of NASA's Astrophysics Data System and of the VALD database operated at Uppsala University. 
\end{acknowledgements}

\bibliographystyle{aa}
\bibliography{bibliography}

\end{document}